\documentclass[runningheads]{svmult}

\usepackage{makeidx}   
\usepackage{graphicx}  
\usepackage{subeqnar}  
\usepackage{multicol}  
\usepackage{physprbb}  
\makeindex             

\newcommand{\gapprox}{\,\rlap{\lower 2.5pt 
\hbox{$\sim$}}\raise 1.5pt\hbox{$>$}\,}
\newcommand{\lapprox}{\,\rlap{\lower 2.5pt 
\hbox{$\sim$}}\raise 1.5pt\hbox{$<$}\,}
\newcommand{\Msun}{M_{\odot}}

\begin{document}
\title*{TP-AGB stars to date high-redshift galaxies with the Spitzer
Space Telescope}
\toctitle{Focusing of a Parallel Beam to Form a Point
\protect\newline in the Particle Deflection Plane}
%
%
\titlerunning{Finding the TP--AGB at redshift above zero}
%
\author{Claudia Maraston
%
\authorrunning{Claudia Maraston}
%
%
\institute{Max Planck Institut f\"ur Extraterrestrische Physik,
Giessenbachstr.\ 1, 85748 Garching b. M\"unchen, Germany}}

\maketitle              

\begin{abstract}
We present new stellar population models that include the contribution
of the Thermally Pulsing Asymptotic Giant Branch (TP-AGB) phase also
in the synthetic spectral energy distribution (SED). The TP-AGB phase
is essential for a correct modeling of intermediate-age ($0.2~\lapprox
t/{\rm Gyr} \lapprox 1\div2$) stellar populations, because it provides
$\sim~40$~\% of the bolometric contribution, and up to $\sim~80$~\% of
that in the $K$-band. These models are obtained by coupling the
energetic of the TP-AGB phase as calibrated with data of Magellanic
Clouds star clusters (\cite{io98}), with empirical spectra of TP-AGB
stars (\cite{LM02}). Now that the Spitzer Space Telescope (SST) allows
the sight of the rest-frame IR at high redshifts, these models provide
the opportunity to use the TP-AGB phase as an age indicator also for
high-redshift stellar populations. Here we focus on redshift $\sim 3$
and provide predictions of the colours of various galaxy models as
will be measured by means of the IRAC imaging instrument on board the
SST. We find a sizable magnitude difference between TP-AGB-dominated
high-redshift stellar populations and those being older or
younger. The first releases of GOODS data should allow a check of
these predictions.
\end{abstract}
\section{The dating of galaxies and the TP-AGB phase}
The epoch(s) of galaxy formation is constrained by dating the stellar
populations at zero as well as at high redshift, because the
timescales of stellar evolution are independent of cosmological
models. Such a constraint provides an important check of current
models of hierarchical galaxy formation, in the framework of which the
assembly of massive galaxies appears to occur over a rather extended
redshift range, with a substantial amount of star formation at
redshift lower than 1 (see reviews by S.~White and R.~Sommerville,
{\it this volume}). Such prediction appears to be at odd with the old
average age, and the small spread in ages, derived for local massive
ellipticals (Es) using optical absorption features and chemical
evolution arguments (Thomas, Maraston, Bender, {\it this volume}; see
also G. Gavazzi, {\it this volume}; \cite{kauffmann+03}). Also, the
finding of galaxies already massive at high redshifts (see
contributions by A. Cimatti; R. Genzel, {\it this volume};
\cite{saraccoetal03}) is difficult to accommodate in such models.
On the other hand, there are also galaxies whose average ages appear
to be rather young and could indeed be consistent with small formation
redshifts ($z\lapprox~1$), like the so-called k+a galaxies
(\cite{biancaetal}), or lenticulars and some low-mass Es in the field
(see Thomas, Maraston, Bender, {\it this volume}).

However, the dating of unresolved stellar populations by means of
spectro-photometric indicators in the optical is limited by the
well-known {\it age/metallicity degeneracy} (\cite{F72}, \cite{W94},
\cite{MT00}), i.e. the phenomenon that a low metallicity can mimick a
low age, and vice versa. When a stellar population ages above $1\div
2$~Gyr, the Red Giant Branch and the Main Sequence share almost
equally the energy production (see e.g. Figure~3 in \cite{io98}), and
their contributions evolve very smoothly with age. At the same time
there are no other stellar phases of short duration and relevant
energetics that become important and could be used as age
indicators. Therefore, at old ages the age/metallicity degeneracy
works at its best in confusing the age determination. To trace back
the beginning of the formation of a stellar system it would be ideal
to recognize it before it becomes a few Gyr old.

A clear signpost of intermediate age ($t\sim 1~\rm Gyr$) stellar
populations are Thermally-Pulsing Asymptotic Giant Branch stars
(TP-AGB; \cite{RB86}, \cite{io98}). According to stellar evolution,
the TP-AGB stellar phase, the brightest and the coolest on the HR
diagram, becomes fully developed in stars with degenerate
carbon-oxygen cores. The onset of such event in the life of a stellar
population has been called the AGB-{\it phase transition}
(\cite{RB86}). The observational evidence of the onset of the TP-AGB
is a sizable jump in the V$-$K colour that increases from
$\sim~1.4$~to $\sim~3.2$, as observed among the Magellanic Clouds
globular clusters (see Section~2). The narrow interval in evolutionary
mass ($1.5 -3~\Msun$) constrains the whole duration of the TP-AGB
dominance to be $\sim$~1 Gyr (\cite{io98}). Therefore, picking up the
TP-AGB is a powerful way of dating a stellar population, and this
technique has been applied with success to local stellar populations
(\cite{ioetal01}).

How can we extend this approach of age dating to $z>0$ ?

An early suggestion in this direction is due to \cite{alvio92}, who
indicated that the AGB phase transition potentially is an effective
tool to date the high redshift formation of Es. Two factors has
hampered the exploitation of this idea until now. First, the
rest-frame IR at redshift $2\div3$ is sampled by the observed frame
around $8-10~\mu$m. This window is only now available thanks to the
advent of the S(pitzer)S(pace)T(elescope). Second, synthetic spectral
energy distributions (SEDs) including the TP-AGB phase are clearly
required, but usually evolutionary population synthesis models include
only the early part of the AGB phase (so-called the Early-AGB),
thereby missing the TP-AGB that is the one energetically important
(see Section~2).

In this contribution we introduce model SEDs that include the TP-AGB
phase (Section~3) and show the substantial effect on the integrated
SEDs of intermediate-age stellar populations. In Section~4 we
construct diagnostic colour-colour diagrams for the imaging instrument
IRAC on board the SST for the illustrative redshift of 3. We further
discuss the use of these models at every redshift.
\section{TP-AGB in SSP models: state of the art}
The TP-AGB is a critical stellar phase to be accounted for in a
Stellar Population (SP) model, because its energetic and duration are
affected by mass-loss and nuclear burning in the envelope, both
phenomena requiring parametrizations to be calibrated with data
(\cite{ir83}). However, the TP-AGB phase is the dominant phase in
intermediate-age stellar populations ($0.2~{\rm Gyr}\lapprox
t\lapprox~1 \div 2~{\rm Gyr}$), contributing $\sim 40\%$ to the
bolometric light, as observed among the globular clusters (GCs) in the
Magellanic Clouds (\cite{FMB90}). Maraston~(1998, \cite{io98}) uses
the {\it fuel consumption} theorem (\cite{RB86}) to include the TP-AGB
phase in SP models in a semi-empirical way, by calibrating the
energetic of the TP-AGB phase with data of intermediate-age Magellanic
Clouds GCs.
\begin{figure}[ht]
\includegraphics[width=.49\textwidth]{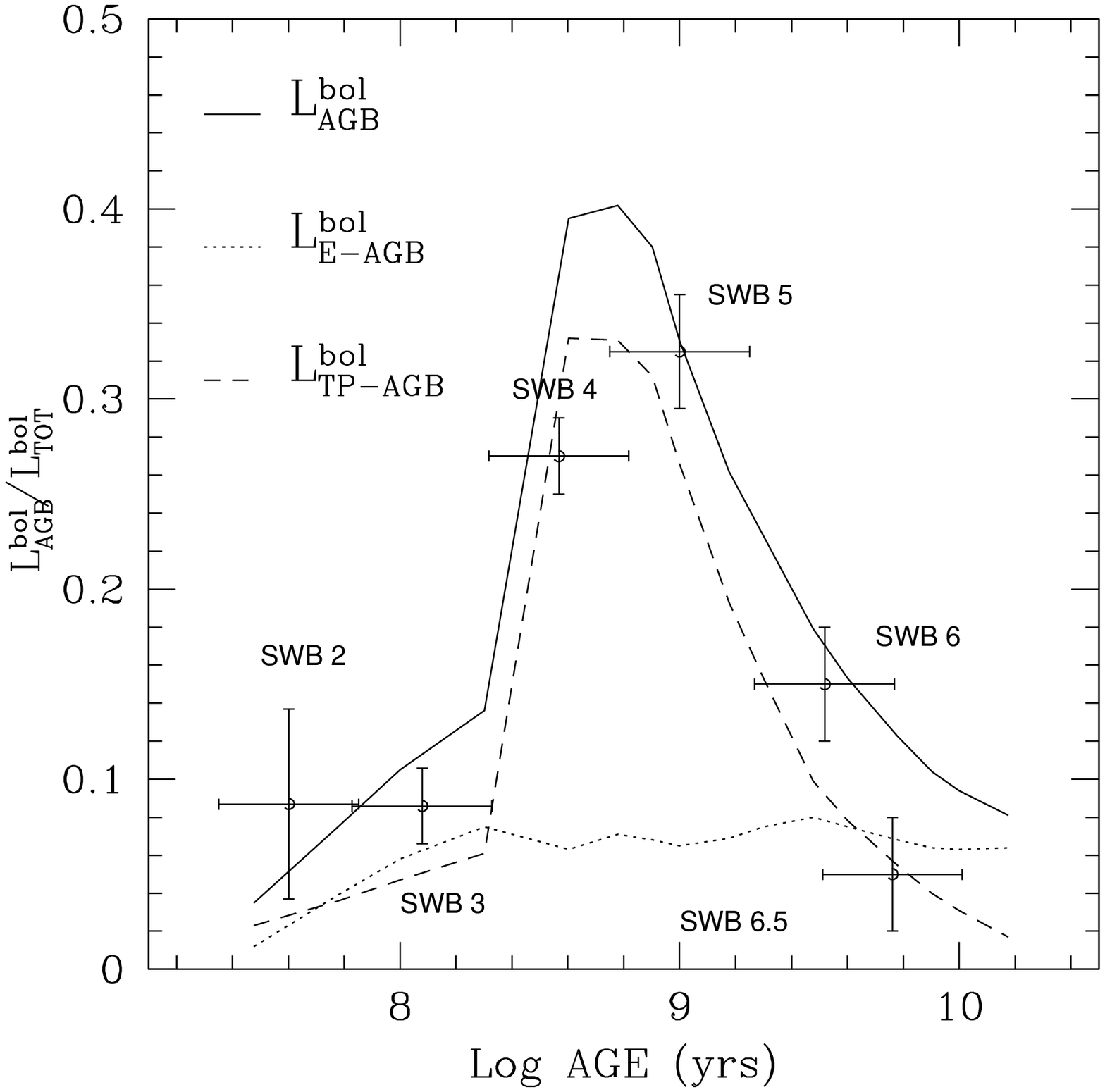}
\includegraphics[width=.49\textwidth]{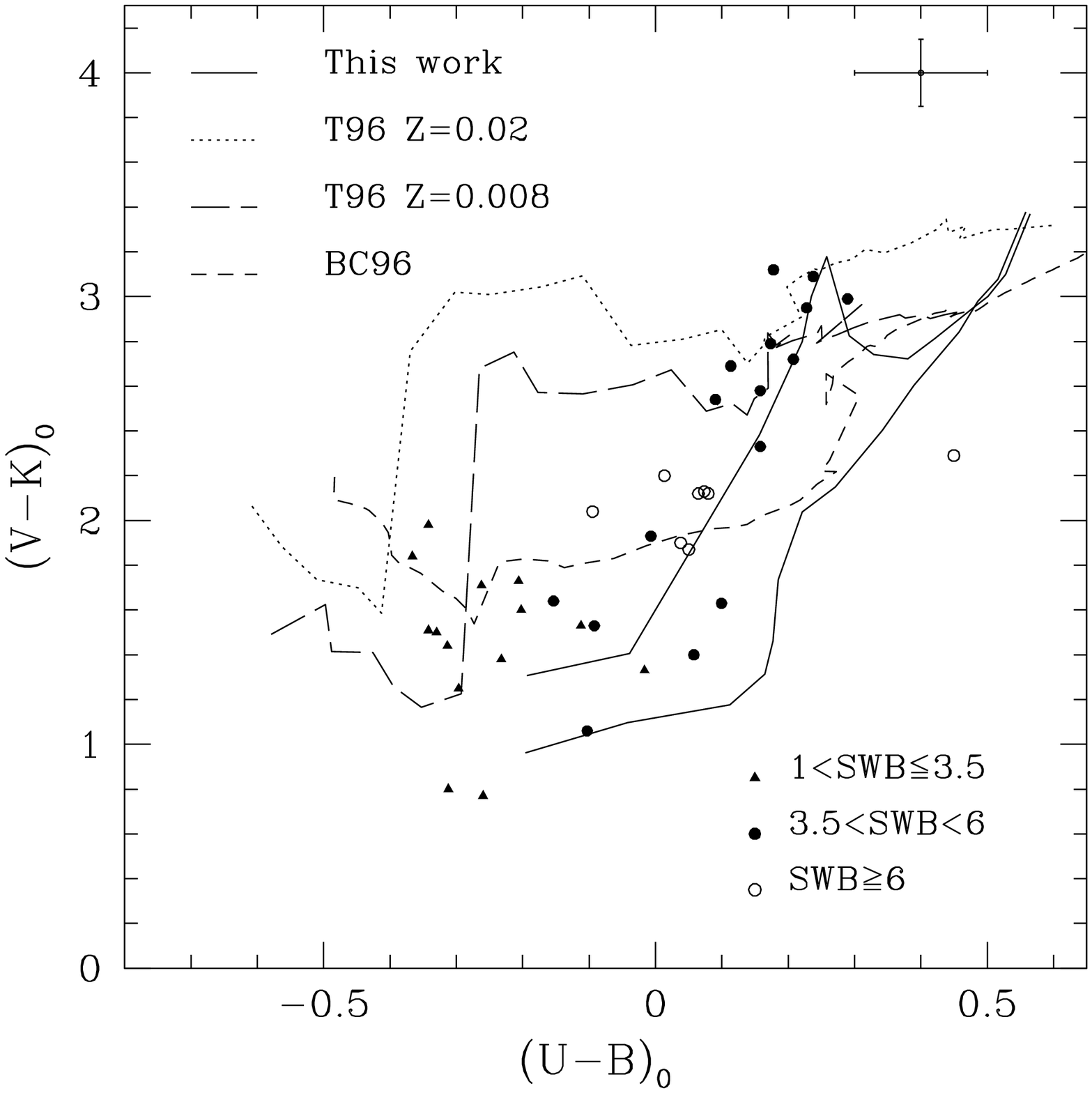}
\caption{From \cite{io98}. {\it Left-hand panel}. Calibration of the
bolometric contribution of the TP-AGB phase of SP models as a function
of age, with data of Magellanic Clouds GCs. {\it Right-hand
panel}. Calibration of the synthetic V$-$K vs. U$-$B (higher solid
line) with the same data. The intermediate-age objects are the filled
circles. The lower solid line shows the same SPs but without the
TP-AGB phase, to appreciate the deficit of IR flux in the latter
case. The other line styles show SPs from other authors (see
\cite{io98}.)}
\label{agb}
\end{figure}
The calibration of the bolometric contribution, and the comparison of
the synthetic broad-band colours with the same data of Magellanic
Clouds GCs, is shown in Figure~\ref{agb}. The AGB phase transition
among the intermediate-age GCs (filled circles) appears as an
enhacement of the IR luminosity with respect to the optical one and
with the V$-$K colour reaching values larger than 3. The figure shows
that the inclusion of the TP-AGB phase (models as thick solid lines)
is crucial to match the integrated near-IR colours of intermediate-age
stellar populations, as emphasized by a model (solid thin line) in
which the TP-AGB contribution has been subtracted on purpouse and only
the E-AGB considered.
The SP models shown in Figure~1 were restricted at the broad-band
colours due to the unavailability at the time of spectra, either
theoretical or empirical, appropriate to TP-AGB stars. However, in
order to use the TP-AGB-induced jump of the near-IR flux as an age
indicator also at $z>0$, this phase has to be included in the
synthetic spectral energy distribution. Such improvements of the
models will be described in the next Section.
\section{TP-AGB in SSP models: extension to SEDs}
Recently, a library of observed spectra of carbon-rich and oxygen-rich
type TP-AGB stars in the wavelength range $0.5~\div~2.5~\mu$m has
become available (\cite{LM02}). We have used these empirical spectra
and the calibrated TP-AGB fuel (\cite{io98}) to include the TP-AGB
phase in the synthetic SEDs of SP models. The SEDs are constrained to
match our previously published model broad-band colours, because these
match the observed colours of the Magellanic Clouds GCs (Figure
\ref{agb}).
\begin{figure}[ht]
\begin{center}
\includegraphics[width=.6\textwidth]{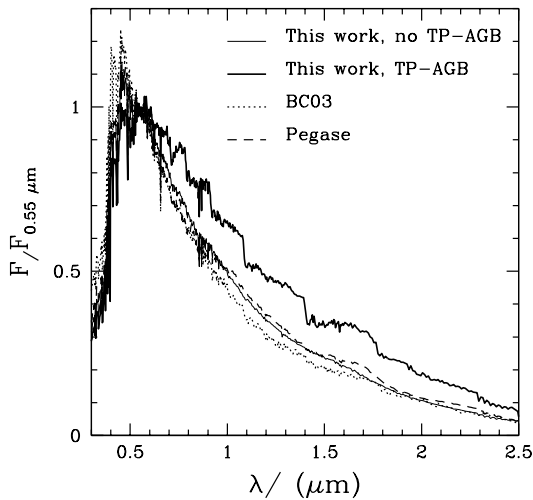}\caption{The effect
of the TP-AGB on the spectral energy distribution of a 1 Gyr old
stellar population. The solid thick and thin lines show our models
with and without TP-AGB. Dotted and dashed lines are models from the
literature (\cite{BC03}, \cite{Pegase}).}
\label{sed}
\end{center}
\end{figure}
Figure \ref{sed} shows as an example the 1 Gyr SED (thick solid
line). The inclusion of the TP-AGB changes substantially the spectrum
at wavelengths $\gapprox 0.7 \mu$m, while leaving unchanged the
optical side. Also shown are the SEDs of other models (\cite{BC03},
\cite{Pegase}. Since the latter do not include the TP-AGB phase in the
synthesis, the detection of intermediate age SP in galaxies based on
the near-IR predictions of these models should be taken with caution.

The models presented here are on course of publication, and we refer
to the article for more details (\cite{io04}). In this proceeding we
use them to compute the observed-frame colours of SP models at
high-redshift in the imaging bands of IRAC-SST to illustrate the power
of the TP-AGB-based diagnostic of SP ages.
\section{How to find the TP-AGB at $z>0$}
At high redshifts when galaxies are dominated by $\sim$~1 Gyr old
populations the TP-AGB signature in the rest-frame near-IR must show
up, with e.g. the rest-frame V$-$K colour mapping into the observed
K$-$10$~\mu$m at $z=3$ (\cite{alvio92}). The conclusion of Chiosi et
al.~(\cite{CBB92}) that the jump in the mid-IR colours due to the AGB
is ``wiped out by cosmological effects'' is due to their consideration
of the {\em observed} V$-$K {\it at every redshift} instead of the
rest-frame one, as appropriate.

The fingerprint of the TP-AGB starts to be appreciable at
rest-wavelength $\sim~0.7\mu$m (Figure \ref{sed}). Already existing
$K$-selected galaxy surveys like MUNICS and the K20 (see Drory;
Cimatti, {\it this volume}) could be used to search for
intermediate-age SP using these models in galaxies at moderate
redshifts ($z\sim1$), where the observed $K$~samples the rest
$J$. This is the subject of a forthcoming paper.

Here we focus on redshift 3 to search for the TP-AGB signature in the
progenitors of massive Es, for which the $\alpha$-enhancement of their
stellar populations constrain the star formation history to be short
(e.g. 1-2 Gyr, see D.~Thomas, {\it this volume}). This maximizes the
relative fraction of stars that spend the TP-AGB phase {\it
simultaneously}. At $z=3$~the rest-frame near-IR maps into the
observed mid-IR ($3~\lapprox~\lambda/\mu m~\lapprox~10$) and therefore
SST is required.
\begin{figure}[ht]
\begin{center}
  \includegraphics[width=\textwidth]{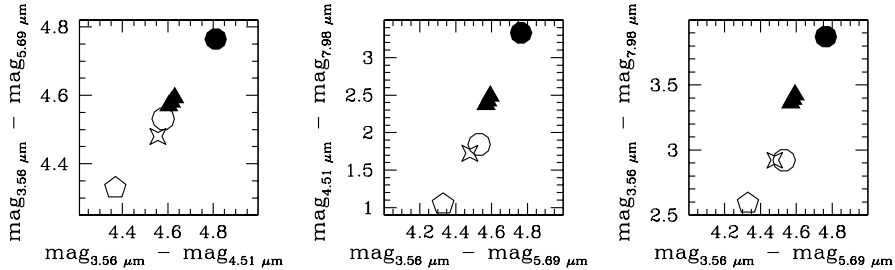}
  \caption{Diagnostic colour-colour diagrams for the IRAC-SST filters
  to detect intermediate-age SP at the illustrative redshift of
  3. Various synthetic stellar populations with solar metallicity are
  shown for which the stars started to form 0.8 Gyr before $z=3$: a
  single burst (SSP, filled circle) and models with exponentially
  declining star formation with e-folding times of 1, 2 and 10 Gyr
  (filled triangles). For comparison, an SSP in which the TP-AGB phase
  is not included is shown as an open circle. Also shown is a younger
  and metal-poor and an older and metal-rich SSP (pentagon and star,
  respectively)}
\label{irac}
\end{center}
\end{figure}
Figure \ref{irac} shows colour-colour diagrams in the four filters
available on the imaging instrument IRAC on board the SST, covering
the region $3.6 \div 8~\mu$m.  The central $\lambda$~of each filter is
indicated in the axis labels. Note that the bluest available IRAC
filter does not map into the rest-frame $V$~at $z=3$, but into the
rest-frame $0.89~\mu$m. While such longer wavelength does not maximize
the signature of the TP-AGB, it has the advantage of being better
sheltered than optical bands from other effects, such as dust
attenuation or newly born hot stars. These effects will be explored
in a forthcoming paper.

Various SP models are shown in Figure~\ref{irac} in which the stars
started to form 0.8 Gyr before redshift 3. The filled circle is a
single burst, or Simple Stellar Population (SSP), that includes the
TP-AGB. The empty circle shows the same model, but in which the TP-AGB
phase has not been included.

Figure \ref{irac} displays the same effect as we saw in Figure
\ref{agb} using the familiar Johnson bands: the presence of TP-AGB
stars enhances the rest-frame mid-IR flux. The single burst maximizes
the effect because of the single age of the stars. However, the TP-AGB
signature is still visible when extended star formation histories are
considered (filled triangles). As mentioned above, this is due to the
fact that at this redshift the three reddest IRAC filters still sample
the mid-IR region, where the impact of the very young stars is
minimized. Also shown in Figure~\ref{irac} is a young, pre TP-AGB
model with lower metallicity (pentagon), and an old, post TP-AGB, but
metal-enriched model (star). As it can be seen, the larger metallicity
is not able to mimick the red colours due to TP-AGB.
From Figure~\ref{irac} we infer a sizable magnitude difference as the
marker of the transition between young and intermediate-age SP that
could be effectively used as age indicator for high-redshift galaxies.
The data collected with SST will allow the TP-AGB-based dating of
stellar populations at large look-back times.

\end{document}